\documentclass{eas}
\usepackage{graphicx}
\usepackage{amsmath}
\usepackage{amssymb}
\usepackage{amstext}
\usepackage{algorithm}
\usepackage{algorithmic}

\begin{document}

\title{Very Efficient Methods for \\
Multilevel Radiative Transfer \\
in Atomic and Molecular Lines} 
\runningtitle{Very Efficient Multilevel Radiative Transfer}
\author{A. Asensio Ramos}
\address{Instituto de Astrofisica de Canarias, 38205, La Laguna, Tenerife, Spain}
\secondaddress{Istituto Nazionale di Astrofisica (INAF), Osservatorio Astrofisico di Arcetri, Largo E. Fermi 5,
50125, Firenze, Italy}
\author{J. Trujillo Bueno}
\sameaddress{1}
\secondaddress{Consejo Superior de Investigaciones Cient\'{\i}ficas, Spain}
\begin{abstract}
The development of fast numerical methods for multilevel radiative transfer (RT) applications often leads to important breakthroughs 
in astrophysics, because
they allow the investigation of problems that could not be properly tackled using the methods previously available. Probably, the
most familiar example is the so-called Multilevel Accelerated $\Lambda$-Iteration (MALI) technique of Rybicki \& Hummer
for the case of a local approximate operator,
which is based on Jacobi iteration. However, there are superior operator-splitting methods, based on Gauss-Seidel (GS) and 
Successive Overrelaxation (SOR) iteration, which provide a dramatic increase in the speed with which non-LTE multilevel transfer problems
 can be solved in one, two and three-dimensional geometries. Such RT methods, which were introduced  by 
Trujillo Bueno \& Fabiani Bendicho ten years ago, are the main subject of
the first part of this paper. We show in some detail how they can be applied for solving multilevel RT problems in spherical geometry, 
for both atomic and molecular line transitions. The second part of the article addresses the issue of the calculation of the molecular 
number densities when the approximation of instantaneous chemical equilibrium turns out to be inadequate, which happens to be the case whenever
the dynamical time scales of the astrophysical plasma under consideration
are much shorter than the time needed by the molecules to form.
\end{abstract}
\maketitle
\section{Introduction}
Astrophysical plasma spectroscopy is a fascinating but very complex research field because non-equilibrium physics is required. 
First, for given atomic or molecular number densities at each point within the astrophysical plasma under consideration, we have the 
problem of how to calculate the population numbers of the levels included in the atomic or molecular model. Second, it is also 
important to remember that the atomic and/or molecular number densities themselves are not always given by the instantaneous values
 of the physical conditions of the plasma (e.g., kinetic temperature, density), simply because such conditions may be
 changing rapidly with time as as result of magneto-hydrodynamical processes. Therefore, for instance, it may be ``dangerous'' to infer chemical abundances in stellar atmospheres by computing 
the molecular number densities via the approximation of instantaneous chemical equilibrium.

The present keynote article addresses these two aspects of astrophysical plasma spectroscopy, with emphasis on the increasingly 
attractive topic of non-LTE radiative transfer in molecular lines (e.g., note that in the very near future the Herschell Space 
Observatory of ESA is expected to be fully operative).

\section{The Non-LTE Multilevel Radiative Transfer Problem}
\subsection{Basic equations}
\label{basic_eq}
The standard multilevel radiative transfer problem requires the joint solution of the radiative transfer (RT) equation (which
describes the radiation field) and the kinetic equations (KE) for the atomic or molecular level populations (which describe the
excitation state). The numerical solution of this non-local and non-linear problem requires to discretize the model atmosphere
in NP points, where the physical properties are assumed to be known. The standard multilevel RT problem consists
in obtaining the population $n_j$ of each of the $j=1,2,\ldots,\mathrm{NL}$ levels included in the atomic/molecular model that are
consistent with the radiation field within the stellar atmosphere. This radiation field has contributions from possible background
sources and from the radiative transitions in the given atomic/molecular model.

Making the usual assumption of statistical equilibrium, the rate equation for each level $i$ at each spatial point reduces to (Socas-Navarro \& Trujillo Bueno 1997):
\begin{equation}
\label{eq_statis_equil_eq}
\sum_{j < i}{\Gamma}_{ji}\,-\,\sum_{j > i}{\Gamma}_{ij}\,
+\,\sum_{j \neq i}{n_j C_{ji}} - n_i \sum_{j \neq i}{C_{ij}} = 0,
\end{equation}
where $C_{ij}$ is the collisional rate between levels $i$ and $j$ and
$\Gamma_{lu}$ is the net radiative rate in the transition from a bound lower level $l$ to an upper level $u$:
\begin{equation}
\label{net_rad_rate}
\Gamma_{lu}\,=\,n_l R_{lu}\,-\,n_u\,R_{ul}\, ,
\end{equation}
with $R_{ij}$ the radiative rates. 
Given that Eqs. (\ref{eq_statis_equil_eq}) are not linearly independent, we replace one of them by the particle conservation law:
\begin{equation}
\label{eq_part_conserv}
\sum_{i} n_i = n_\mathrm{total},
\end{equation}
where $n_\mathrm{total}$ is the total number density of the species under consideration.
The resulting system can be formally written as:
\begin{equation}
\label{eq_matricial_sys}
\bf A \cdot n = f,
\end{equation}
where $\bf A$ is a matrix of size NL$\times$NL whose elements contain the collisional and radiative rates (except for one of the rows that contains 1 due to the conservation law), $\bf f$ is a vector of length NL with zeros
except for the $n_{\mathrm{total}}$ value of the conservation law, and $\bf n$ is a vector containing the population of each level.
The result of grouping together the sets of Eqs. (\ref{eq_matricial_sys}) for all the NP grid points of the atmosphere
can be symbolically represented as
\begin{equation}
\label{eq_matricial_sys_total}
\bf L \cdot n = f,
\end{equation}
where $\mathbf{f}$ is a known vector of size NP$\times$NL, $\mathbf{n}$ is a vector of the same length with the populations of
the NL levels for each of the NP points and $\mathbf{L}$ is
a \emph{block-diagonal} matrix such that each of its NP blocks of size NL$\times$NL is given by the matrix $\mathbf{A}$ of Eq.
(\ref{eq_matricial_sys}).

The collisional rates are assumed to be known and given by the local physical conditions of the atmosphere. On the other
hand, the net radiative rates $\Gamma_{lu}$ depend on the radiation field within the stellar atmosphere. For bound-bound transitions
\begin{equation}
\label{net_rate_bound}
\Gamma_{lu}\,=(n_lB_{lu}-n_uB_{ul}){\bar{J}}_{lu}\,-\,n_uA_{ul},
\end{equation}
where $A_{ul}$, $B_{ul}$ and $B_{lu}$ are the Einstein coefficients. A key quantity is $\bar J_{lu}$, which
is the frequency averaged mean intensity weighted by the line absorption profile:
\begin{equation}
\label{eq_mean_intensity}
\bar J_{lu} = \frac{1}{4 \pi} \int d \mathbf{\Omega} \int d \nu \phi_{lu} (\nu,\mathbf{\Omega})
I_{\nu \mathbf{\Omega}},
\end{equation}
where $\phi_{lu}$ and $I_{\nu \bf \Omega}$ are, respectively, the normalized line profile and
the specific intensity at frequency $\nu$ and direction $\bf \Omega \rm$. The specific intensity is governed by the radiative 
transfer equation, which can be formally solved if we know the variation of the opacity ($\chi_{\nu}$) and of the source function ($\epsilon_{\nu}/\chi_{\nu}$, being $\epsilon_{\nu}$ the emission coefficient)
in the medium. Once the stellar atmosphere is discretized, the specific intensity can be written formally as
\begin{equation}
\label{eq_formal_sol}
\bf I_{\nu \mathbf{\Omega}} = \bf \Lambda_{\nu \mathbf{\Omega}} \left[ \bf S_{\nu} \right] + \mathcal{\bf T}_{\nu \mathbf{\Omega}}, \nonumber
\end{equation}
where $\mathcal{\bf T}_{\nu \mathbf{\Omega}}$ is a vector that accounts for the contribution of the boundary conditions to the intensity
at each spatial point of the discretized medium, $\bf S_{\nu}$ is the source function vector 
and $\bf \Lambda_{\nu \mathbf{\Omega}}$ is an operator whose element $\Lambda_{\nu\mathbf{\Omega}}(i,j)$ gives the response of the radiation field at point ``$i$'' due to a unit-pulse perturbation in the source function at point ``$j$''.

Since the radiative transfer equation couples different parts of the atmosphere and the absorption and emission properties at all the spatial points depend on the level populations, the RT problem is both \emph{non-local} and \emph{non-linear}. Therefore, the system of Eqs. (\ref{eq_matricial_sys}) represents a highly non-linear problem (i.e., the operator $\mathbf{A}$ depends implicitly on $\mathbf{n}$). This non-linearity makes it necessary to apply suitable iterative methods.

\subsection{Formal solution in spherical geometry}
\label{sec_formal_solution}
The equation that describes the radiative transfer in a spherically symmetric atmosphere where the physical conditions vary only along the radial direction is (e.g., Mihalas 1978):
\begin{equation}
\label{eq_RT_eq_spher}
\mu \frac{\partial I_{\nu}(r,\mu)}{\partial r} + \frac{\left( 1 - \mu^2 \right)}{r}
\frac{\partial I_{\nu}(r,\mu)}{\partial \mu} = \epsilon_{\nu}(r,\mu) - \chi_{\nu}(r,\mu) I_{\nu}(r,\mu),
\end{equation}
where $r$ is the radial coordinate and $\mu$ the cosine of the angle between the radial
direction and the ray. This partial differential 
equation (PDE) can be reduced to a set of ordinary differential equations (ODE) if solved along the
characteristics of the PDE. In this case, the characteristic curves (parameterized by $t$) are straight lines with constant impact parameter $p$:
\begin{eqnarray}
\frac{\partial r}{\partial t} &=& \mu \nonumber \\
\frac{\partial \mu}{\partial t} &=& \frac{1-\mu^2}{r}.
\end{eqnarray}

In order to solve the RT equation in spherical geometry, we apply the parabolic short-characteristics (SC) method 
(Kunasz \& Auer 1988; Auer \& Paletou 1994; Auer, Fabiani Bendicho \& Trujillo Bueno 1994) along each
ray of constant impact parameter. As shown by Trujillo Bueno \& Fabiani Bendicho (1995), the SC method allows an efficient implementation of the fast iterative methods based on the Gauss--Seidel iterative scheme. The parabolic SC method assumes that the
variation of the source function with the optical depth 
along the ray under consideration has a parabolic behavior
between each three consecutive points. Consider the \emph{upwind} point M, the point O where the intensity is being calculated and the \emph{downwind} point P. The source function at each of such consecutive points can be easily calculated from the current 
values of the level populations. The intensity $I_\mathrm{M}$ at the upwind point M is also known from previous steps. Taking into 
account that the formal solution of the radiative transfer equation is given by
\begin{equation}
I_\mathrm{O} = I_\mathrm{M} e^{-\Delta \tau_{\mathrm{MO}}} + \int_0^{\Delta \tau_{\mathrm{MO}}} S(t) 
e^{-(\Delta \tau_{\mathrm{MO}}-t)} dt , 
\end{equation}
it is easy to obtain the following formula (Kunasz \& Auer 1988):
\begin{equation}
\label{eq_short_characteristics}
I_\mathrm{O} = I_\mathrm{M} e^{-\Delta \tau_\mathrm{MO}} + \Psi_\mathrm{M} S_\mathrm{M} + \Psi_\mathrm{O} S_\mathrm{O} + \Psi_\mathrm{P} S_\mathrm{P}.
\end{equation}
The quantities $\Psi_\mathrm{M,O,P}$ are functions of the optical distances $\Delta \tau_\mathrm{MO}$ between the upwind point
M and the point O and of the optical distance $\Delta \tau_\mathrm{OP}$ between the point O and the downwind
point P, while $S_\mathrm{M,O,P}$ are the values of the source function at points M, O and
P, respectively. We can build a similar formula assuming that the source function varies linearly between points M and O, which is used only at the boundary grid points for rays going out of the boundaries. In this case,
the last term of Eq. (\ref{eq_short_characteristics})
disappears and $\Psi_\mathrm{M}$ and $\Psi_\mathrm{O}$ only depend on $\Delta \tau_\mathrm{MO}$.
Note that $\Delta \tau_\mathrm{MO}{\approx}(\chi_\mathrm{M}+\chi_\mathrm{O})/2$ and
$\Delta \tau_\mathrm{OP}{\approx}(\chi_\mathrm{O}+\chi_\mathrm{P})/2$, with $\chi_i$ the opacity at point $i$ (which depends on the lower-level population at point $i$).

\subsection{Iterative methods for radiative transfer applications}
\label{sec_iterative_methods}
The basic idea behind an iterative method for the solution of a nonlinear problem is to start from an estimation of the
solution (ideally as close to the final solution as possible) and then to perform successive corrections to the current estimation until the final solution is found. Iterative methods for multilevel transfer problems obey the same structure. Consider the solution of Eqs. (\ref{eq_matricial_sys}) at each of the NP spatial
points. We start from an
estimate, $\bf n^{\rm old}$, of the atomic or molecular level populations at each point.
If this estimate is not the exact solution of the problem, we have a nonzero residual
\begin{equation}
\label{eq_residual}
\bf r = f - A^{\rm old} \cdot n^{\rm old} \neq 0,
\end{equation}
where the superscript ``old'' indicates that the radiative rates of the operator $\bf A^{\rm old}$ have to be calculated by solving the
RT equation using the previous estimation of the
populations (i.e., using $\bf n^{\rm old}$). This matrix is built, for every point in the atmosphere, by using also the appropriate collisional rates of the chosen atomic/molecular model. The aim is to find the correction, $\mathbf{\delta n}$, such that
the estimation:
\begin{equation}
\label{eq_correction}
\bf n^{\rm new} = n^{\rm old} + \delta n
\end{equation}
gives a zero residual and, therefore,
\begin{equation}
\label{eq_final_sol}
\bf A^{\rm new} \cdot \delta n = f - A^{\rm new} \cdot n^{\rm old}.
\end{equation}
Note that this equation cannot be directly solved because it is still nonlinear and similar to the original Eq.
(\ref{eq_matricial_sys}); hence it is not possible to solve the problem in one single step. Therefore, we have to obtain approximate corrections in order 
to arrive to the self-consistent solution iteratively.
The different iterative methods applied to the solution of multilevel radiative transfer problems differ in the way one manages to build, at each iterative step, a linear
system of equations whose solution gives \emph{approximate} corrections to the level populations. An efficient iterative method should account for the non-locality and the radiative couplings of the problem (see Socas-Navarro \& Trujillo Bueno 1997). 

\subsubsection{The $\Lambda$-iteration method}
\label{eq_lambda_iter}
This method is the most direct and simple one. It consists in solving Eq. (\ref{eq_final_sol}) by calculating the radiative rates at each point in the atmosphere using the populations from the previous iterative step (i.e., $\bf n^{\rm old}$). Therefore, the corrections to the estimation of the populations can be obtained from:
\begin{equation}
\label{eq_lambda_iter_scheme}
\bf A^{\rm old} \cdot \delta n = r.
\end{equation}
Note that the mean intensity $\bar J_{lu}$ at each spatial point is obtained by solving the RT equation using the
$\mathbf{n}^{\mathrm{old}}$ values of the populations at all the points of the computational grid. In this way, at each 
iterative step the nonlinear system
is transformed into a linear one, which can be
solved easily to obtain the approximate corrections to the populations. This procedure is iterated until convergence. Although very
simple and easy to code, the $\Lambda$-iteration method has a serious drawback,
which is its very poor convergence rate in optically thick media. This is due to the fact that this method is equivalent to assuming that the radiation field does not react
to the perturbations in the populations (see Socas-Navarro \& Trujillo Bueno 1997). For example, for the case of bound-bound transitions 
such an unrealistic assumption implies that the
expression of Eq. (\ref{net_rate_bound}) is taken equal to 
\begin{equation}
\label{net_rate_lambda}
\Gamma_{lu}\,=(n_l^{\rm new}B_{lu}-n_u^{\rm new}B_{ul}){\bar{J}}_{lu}^{\rm old}\,-\,n_u^{\rm new}A_{ul}.
\end{equation}
In fact, when this approximate expression for the net radiative rate is introduced in Eq. (\ref{eq_statis_equil_eq}) we obtain Eq. (\ref{eq_lambda_iter_scheme}). In any case, we point out that
the $\Lambda$-iteration method is useful for
optically thin problems (e.g., Bernes 1979, Uitenbroek 2000).

\subsubsection{The Multilevel Accelerated $\Lambda$-iteration Method (MALI)}
\label{sec_accel_lambda_iter}
MALI is a clever modification of the $\Lambda$-iteration method, which implies a much higher convergence
rate (Cannon 1973; Olson, Auer \& Buchler 1986; Rybicki \& Hummer 1991, 1992; Auer, Fabiani Bendicho \& Trujillo Bueno 1994).
There are several methods to achieve the required linearity in the system of Eqs. (\ref{eq_final_sol}) at each iterative step. The most powerful ones are linearization
(Auer \& Mihalas 1969; Scharmer \& Carlsson 1985)
and preconditioning (Rybicki \& Hummer 1991, 1992). Socas-Navarro \& Trujillo Bueno (1997) showed that both schemes
are equivalent, but that the preconditioning
strategy with a local approximate operator guarantees positive populations. 
The preconditioning approach with a local approximate operator is nowadays the most used due to its simplicity and good performance. It is usually referred to as Multilevel-ALI (MALI).
 
As pointed out by Socas-Navarro \& Trujillo Bueno (1997) this method takes into account the local response of the radiation field to the source function perturbations. For example, for the case of bound-bound transitions the 
expression of Eq. (\ref{net_rate_bound}) takes the form
\begin{equation}
\label{net_rate_mali}
\Gamma_{lu}\,=(n_l^{\rm new}B_{lu}-n_u^{\rm new}B_{ul})({\bar{J}}_{lu}^{\rm old}-{\bar{\Lambda}}_{lu}^{*}S_{lu}^{\rm old})\,-\,n_u^{\rm new}A_{ul}(1-{\bar{\Lambda}}_{lu}^{*}),
\end{equation}
where $S_{lu}^{\mathrm{old}}$ is the line source function computed using $\mathbf{n}^{\mathrm{old}}$. When this new approximate expression for the net radiative rate is introduced in Eq. (\ref{eq_statis_equil_eq}) the resulting linearized system of equations is 
\begin{equation}
\label{eq_MALI_scheme}
\bf A_{\rm MALI}^{\rm old} \cdot \delta n = r,
\end{equation}
which is similar to Eq. (\ref{eq_lambda_iter_scheme}), but with
the crucial difference that the elements of $\bf A_{\rm MALI}^{\rm old}$ now account for the information provided by the approximate $\Lambda$-operator through the quantity
\begin{equation}
\bar{\mathbf{\Lambda}}_{lu}^* = \frac{1}{4\pi} \int d\mathbf{\Omega} \int d\nu \phi_{lu}(\nu) \mathbf{\Lambda}_{\nu \mathbf{\Omega}}^{*}
r_{lu}^{\mathrm{old}},
\end{equation}
where $r_{lu}^{\mathrm{old}}$ is the frequency-dependent line strength given by
\begin{equation}
r_{lu}^\mathrm{old} = \frac{\chi_{lu}^\mathrm{old} \phi_{lu}(\nu)}{\chi_{lu}^\mathrm{old} \phi_{lu}(\nu) + \chi_c},
\end{equation}
with $\chi_{lu}^\mathrm{old}$ the line opacity computed with $\mathbf{n}^\mathrm{old}$ and $\chi_c$ the background opacity.
When the approximate operator is chosen equal to the diagonal 
of the full operator $\bf \Lambda_{\nu \mathbf{\Omega}}$, the computational time per iteration is similar to that of the $\Lambda$-iteration scheme. 

The implementation of the MALI method for the solution of RT problems in spherical geometry can be easily carried out with
the aid of the short-characteristics formal solver. We need to evaluate the mean intensity, $\bar J_{lu}$, and the approximate operator, 
$\bar{\mathbf{\Lambda}}_{lu}^*$, for each transition. The following steps describe the implementation details:
\begin{enumerate}
\item[1] \emph{Incoming part}. We start the integration of the RT equation from the outer boundary surface along rays with constant impact parameter $p$ using the appropriate boundary condition. 
If the outer boundary condition has a dependence on the angle $\mu$, then each characteristics
will have a different boundary condition $I(R,\mu)$. The intensity is then propagated inwards along each ray applying the formal solution
as established by the parabolic short-characteristics method. This process is carried out for each ray
until the inner boundary surface is reached (for the rays that intersect the core) or until tangentially reaching the shell with $r=p$ (for
the rays that do not intersect the core). Such calculations give us the information required to obtain the angular dependence of the
incoming radiation field and its contribution to the mean intensity:
\begin{equation}
\bar J_{lu}(\mathrm{in},r_i) = \frac{1}{2} \int_{-1}^{0} d\mu \int d\nu \phi_{lu}(\nu) I(r_i,\mu),
\label{eq_mean_intensity_in_MALI}
\end{equation}
where we have explicitly indicated that it is the contribution of the incoming radiation. The contribution of the incoming radiation to the approximate $\mathbf{\Lambda}_{\nu \mathbf{\Omega}}^*$ operator is also calculated. Since for this operator we have
chosen the diagonal of the
full operator, it represents the response
of the mean intensity to a unit pulse perturbation in the source function at the point O under consideration. For the case of a bound-bound
transition with a fixed background continuum, it can be calculated as:
\begin{equation}
\bar{\Lambda}^*_{ii}(\mathrm{in},r_i) = \frac{1}{2} \int_{-1}^{0} d\mu \int d\nu \phi_{lu}(\nu) \left[ I_\mathrm{M} 
e^{-\Delta \tau_\mathrm{MO}^{\mathrm{old}}} + \Psi_\mathrm{O}^{\mathrm{old}}(\mathrm{in})\right] 
r_{lu}^\mathrm{old} ,
\end{equation}
where $\Psi_\mathrm{O}$ is one of the coefficients of the short-characteristics formula (\ref{eq_short_characteristics}). Although 
$I_\mathrm{M}$ is usually taken equal to zero, in reality it is a small negative contribution which can be easily calculated (see the shaded area of Fig. \ref{fig_source_IM}). 
The reason is that the unit pulse perturbation at point O produces a residual intensity at point M coming from the application of the
parabolic SC formula to the points L, M and O, as indicated in Fig. \ref{fig_source_IM}.

\begin{figure}
\centering
\includegraphics[width=8cm]{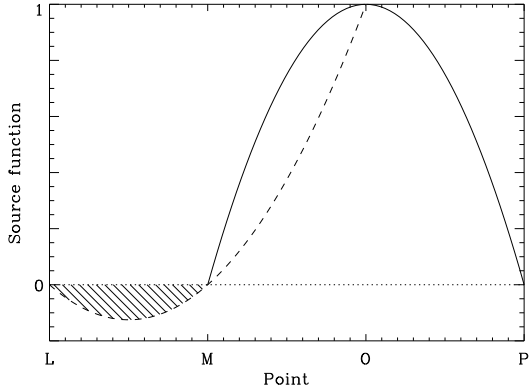}
\caption{The unit pulse perturbation in the source function at point O necessary for the calculation of the 
$\mathbf{\Lambda}_{\nu \mathbf{\Omega}}^*$ operator
produces a residual intensity at point M. This contribution comes from the application of the parabolic SC formula at points L, M and O, where
point L precedes point M along the ray.}
\label{fig_source_IM}
\end{figure}

\item[2] \emph{Outgoing part}. We apply the inner boundary condition for the rays that intersect the core and continue integrating the RT equation
along the outgoing rays until reaching the external boundary surface. The contribution of the outgoing radiation field to the mean
intensity can be obtained as:
\begin{equation}
\bar J_{lu}(\mathrm{out},r_i) = \frac{1}{2} \int_{0}^{1} d\mu \int d\nu \phi_{lu}(\nu) I(r_i,\mu),
\end{equation}
and similarly for the approximate operator.
\end{enumerate}
When this process is carried out for all the spectral lines of the atomic/molecular model, the mean intensity $\bar J_{lu}$ and the  
$\bar{\mathbf{\Lambda}}_{lu}^*$ operator are available at each spatial point. We can then build the system matrix $\bf A_{\rm MALI}^{\rm old}$ for
each point and obtain the approximate corrections to the populations. 

\subsubsection{The Multilevel Gauss--Seidel Method (MUGA)}
\label{eq_mult_gauss_seidel}
This method was developed by Trujillo Bueno \& Fabiani Bendicho (1995). The paper by Fabiani Bendicho, Trujillo Bueno \& Auer (1997) gives a suitable summary of its application to the multilevel problem in cartesian coordinates, including a detailed analysis of its convergence rate and performance. 
Although the computational
time per iteration is similar to that of MALI, MUGA has a much better convergence behavior. 
The key point of the Gauss--Seidel method for RT applications is that one obtains the convergence rate of an upper or lower
triangular approximate $\Lambda$-operator without having to build and/or invert this triangular operator. To this end, once the radiation field is
known at the atmospheric point being considered, the population correction can be made directly using Eqs. (\ref{eq_MALI_scheme}). Then,
if these new populations
are taken into account when calculating the radiation field at the next spatial grid point, the resulting scheme turns out to be equivalent
to that of the Gauss-Seidel method, which has a higher convergence rate than Jacobi's method 
(Trujillo Bueno \& Fabiani Bendicho 1995; Trujillo Bueno 2003). It is important to clarify that MUGA, contrary to what happens
with MALI, requires an specific order of the loops over transitions, angles (equivalently, rays with constant impact parameter in the spherically 
symmetric case) and spatial points when propagating the radiation. \emph{The most external loop is the one over directions, first going
inwards and then ouwards. The next one is the loop over spatial points, followed by the loop over transitions. The most internal ones are the 
loops over angles and frequencies.}
The reason is that we need to evaluate the mean
intensity for \emph{all} the radiative transitions in order to be able to perform the population correction before advancing to the following point 
of the spatial grid. This ordering is shown in a pseudo-code manner in the following algorithm:

\begin{algorithm}
\caption{The MUGA scheme}
\label{alg_1}
\begin{algorithmic}
\FOR[directions are \emph{incoming} and \emph{outgoing}]{directions}
\FOR{spatial points}
\FOR{transitions}
\FOR{angles}
\FOR{frequencies}
\STATE Propagate radiation using parabolic SC
\ENDFOR
\ENDFOR
\STATE Calculate mean intensity of the given transition
\ENDFOR
\STATE Carry out population correction
\ENDFOR
\ENDFOR
\end{algorithmic}
\end{algorithm}

\begin{figure}
\centering
\includegraphics[width=5cm]{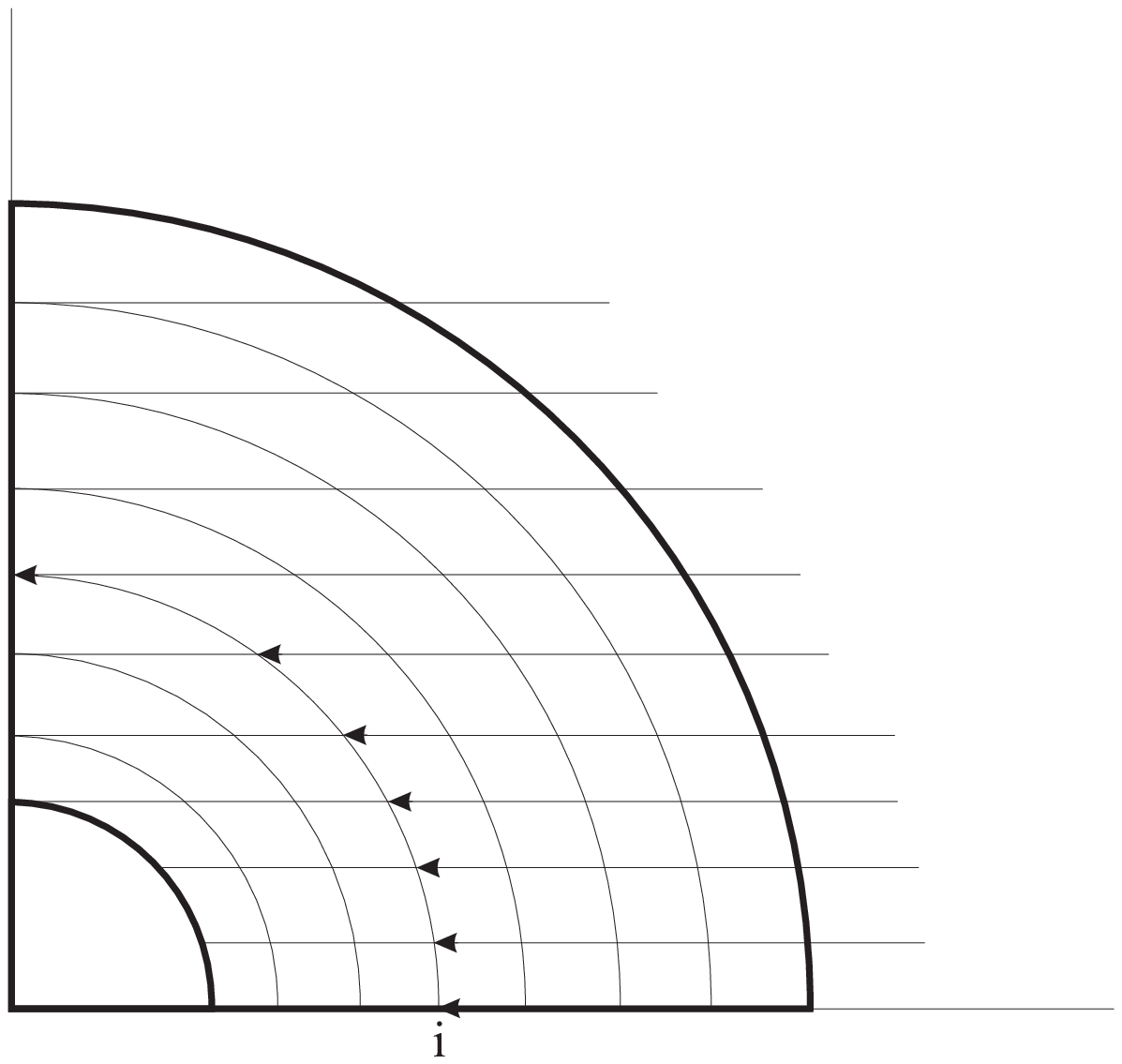}\hspace{1cm}%
\includegraphics[width=5cm]{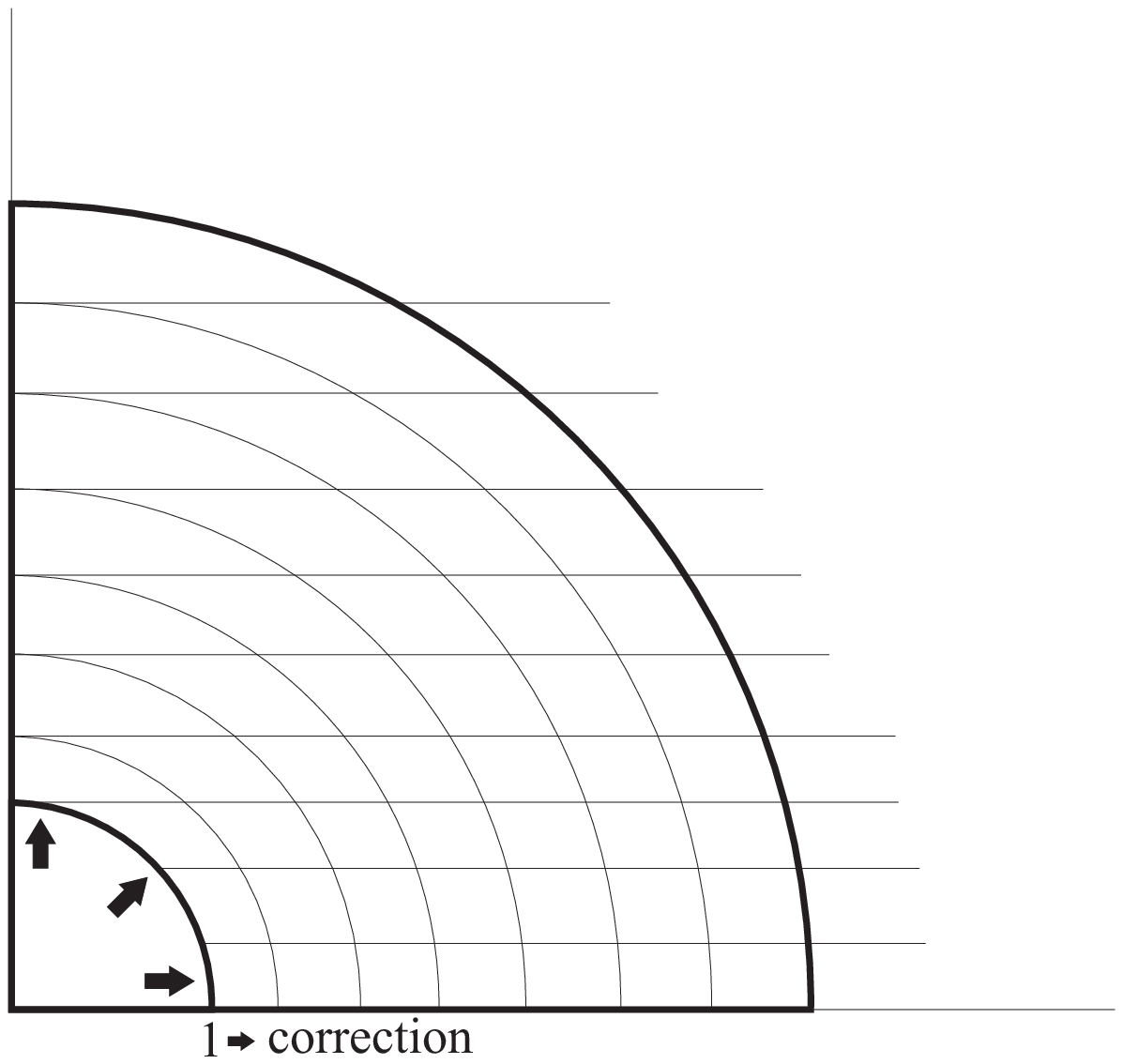}
\includegraphics[width=5cm]{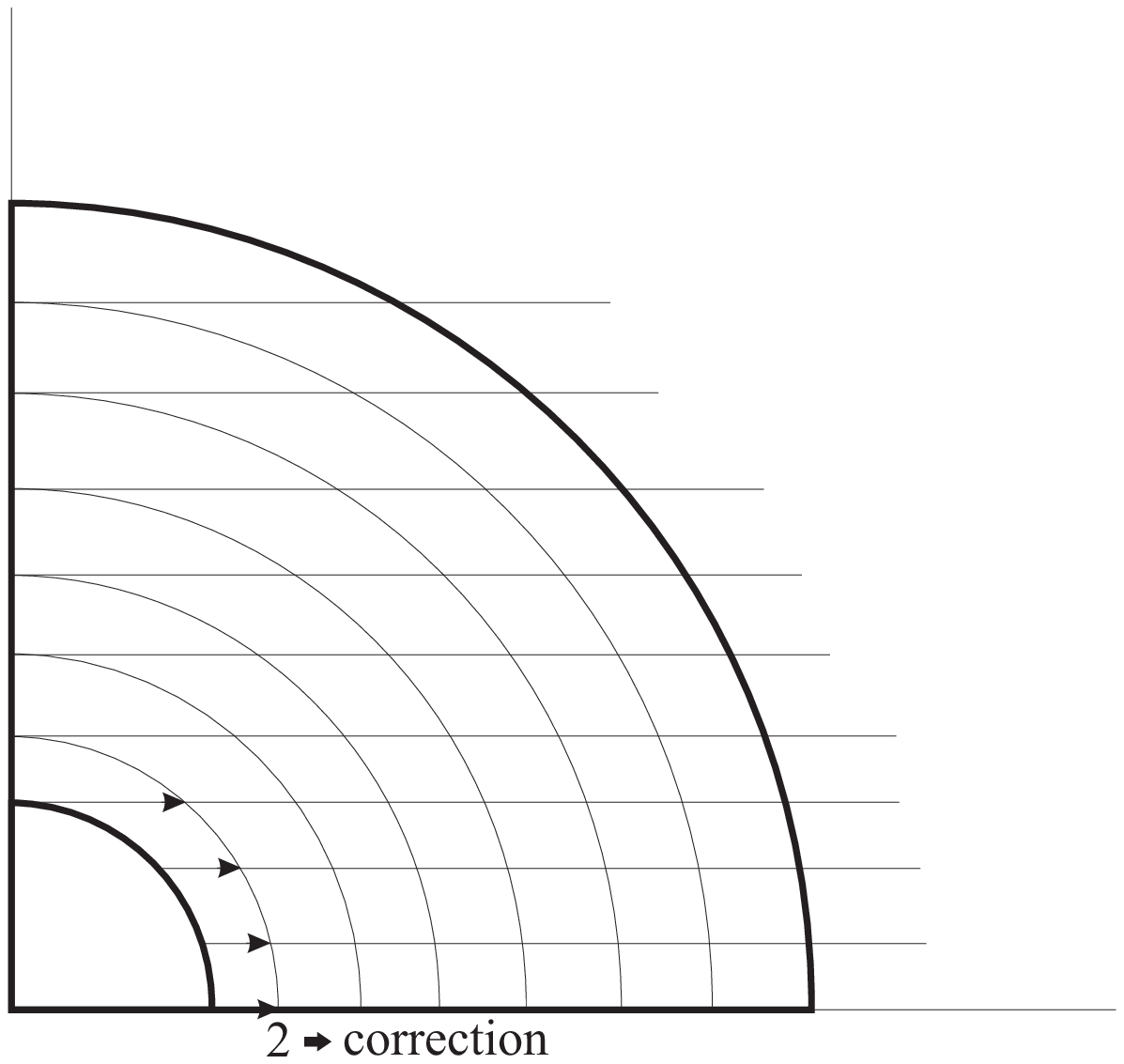}\hspace{1cm}%
\includegraphics[width=5cm]{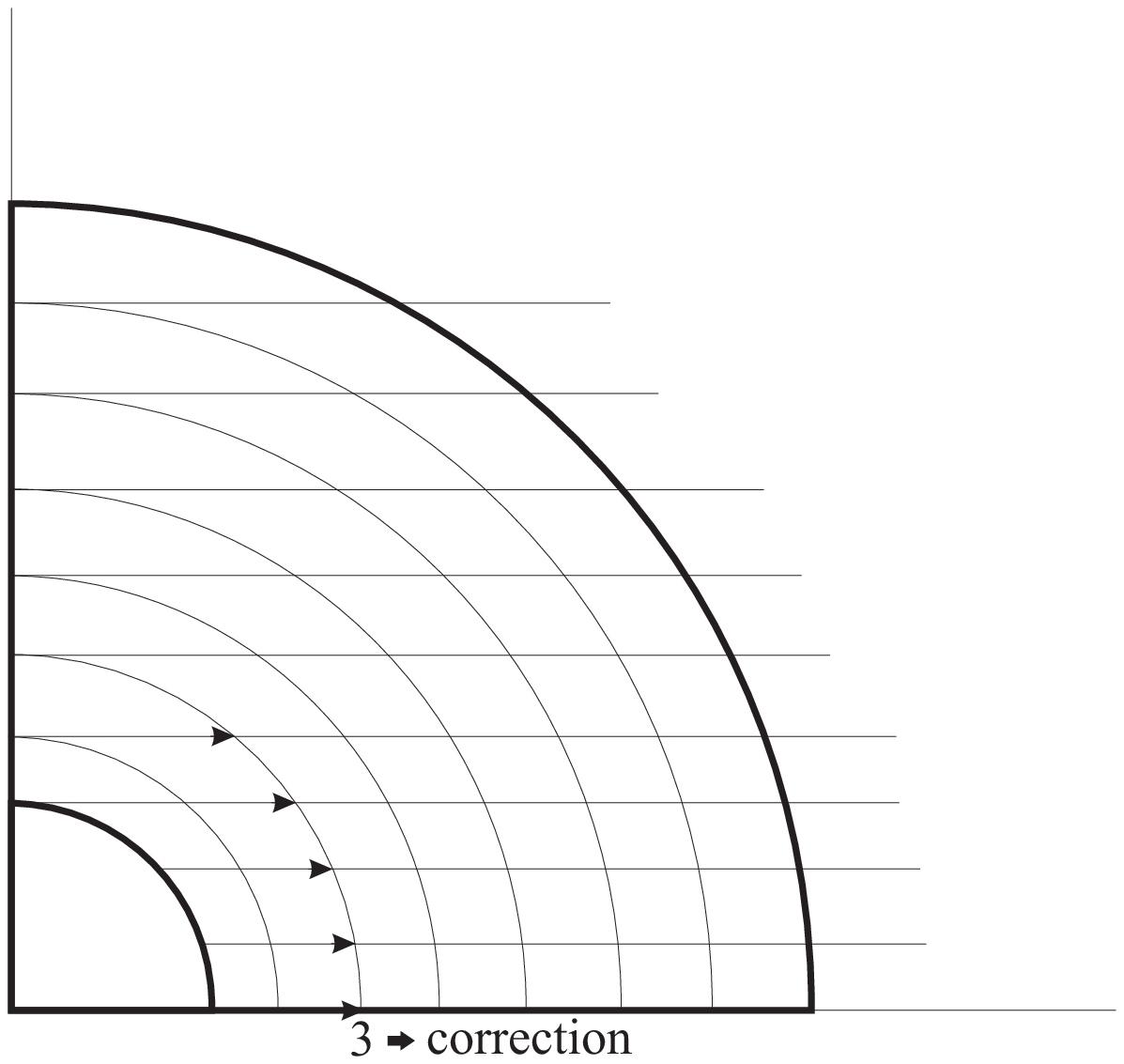}
\caption{Schematic representation of the Gauss-Seidel iterative scheme in its incoming and outgoing phases. Note that the inner 
boundary is the shell $i=1$, while the external boundary is the shell $i=\mathrm{NP}$.
}
\label{fig_GS_spherical_scheme}
\end{figure}

MUGA can be extended to spherical geometry as follows:
\begin{enumerate}

\item[1] \emph{Incoming part}. The formal integration of the RT equation along each ray of constant impact parameter $p$
is started at the outer boundary. This process is schematized in the upper left panel of
Fig. \ref{fig_GS_spherical_scheme}. Note that the intensity can be propagated simultaneously along
all the rays. The integration along each ray is done using short-characteristics,
calculating the intensity at each shell until the ray
intersects the central core or arrives tangent to the shell $r = p$.
Once the incoming intensity at every intersection point between the shells and the whole set of
rays is known, the contribution to the mean intensity of the incoming rays ($\mu<0$) can be calculated.

\item[2] \emph{Outgoing part}. When the contribution to the mean intensity from the incoming radiation is
known, the boundary condition at the inner core can be applied to get the
contribution of the outgoing radiation to the mean intensity in all the radiative transitions, thus allowing us to solve the
linear system of Eqs. (\ref{eq_MALI_scheme}) to get the population correction at
the inner core. The schematic representation of this step is shown in the upper right panel of Fig. \ref{fig_GS_spherical_scheme}.
With these improved level populations, one can propagate the intensity outwards
until the next shell and get the mean intensity there for each of the transitions [what Trujillo Bueno \& Fabiani Bendicho (1995) call $\bar J_{lu}^{\mathrm{old \& new}}$], which then
allows us to solve the linear system of Eqs. (\ref{eq_MALI_scheme}) to obtain the new populations corresponding
to this shell. The procedure is
repeated until arriving to the outer boundary. This process is schematized in the lower left and right panels of Fig.
\ref{fig_GS_spherical_scheme}. The ``old\&new'' label of the mean intensity is used to indicate
clearly that the value of the mean intensity $\bar J_{lu}$ at shell $n$ is obtained by using the new population at the shells with
$r<r_n$,
contrarily to what happens with the MALI method, where the mean intensity is calculated always by using the ``old'' population estimation.
Consequently, the final system to be solved at each shell is:
\begin{equation}
\label{eq_GS_scheme}
\bf A_{\rm GS}^{\rm old \& new} \delta n = r,
\end{equation}
where the coefficients of $\bf A_{\rm GS}^{\rm old \& new}$ are similar to those of $\bf A_{\rm MALI}$ but calculated one
spatial point after the other as summarized above.

\end{enumerate}

As explained in detail by Trujillo Bueno \& Fabiani Bendicho (1995) for the plane--parallel case,
two corrections have to be performed 
in order to obtain a true
GS iteration. Let us assume that we have already obtained the new populations at point $i$ and that we want to calculate the 
new populations at point
$i+1$. The corrections, graphically schematized in Figure \ref{fig_GS_corrections}, are the following:

\begin{enumerate}
\item[a)] The first one, represented in the left panel of Fig. \ref{fig_GS_corrections}, affects the incoming radiation field at point $i+1$ after
the population correction is performed at point $i$ during the outgoing pass. Since the parabolic SC formal solver is being used, the radiation field at
point $i+1$ depends on the absorption and emission properties at points $i$, $i+1$ and $i+2$. The incoming radiation field at $i+1$ had been
calculated using the ``old'' populations at point $i$. Since we have already a new estimation of the populations at point $i$, we have to
correct the
incoming radiation field as illustrated in Fig \ref{fig_GS_corrections}. We indicate in grey those points for which we have a ``new'' estimation of the
populations, while those points with the ``old'' populations are marked in black.
The intensity at point $i+1$ for the incoming ray with impact parameter $p$ was calculated according
to the formula:
\begin{equation}
\begin{split}
I_{i+1}^{\mathrm{old}}(\mathrm{in},p) &= I_{i+2}^{\mathrm{old}}(\mathrm{in},p) e^{-\Delta \tau_{i+2,i+1}^{\mathrm{old}}}  \\
&+ \Psi_{i+2}^{\mathrm{old}}(\mathrm{in}) S_{i+2}^{\mathrm{old}}(p) +\Psi_{i+1}^{\mathrm{old}}(\mathrm{in}) S_{i+1}^{\mathrm{old}}(p) + \Psi_{i}^{\mathrm{old}}(\mathrm{in}) S_{i}^{\mathrm{old}}(p),
\end{split}
\end{equation}
which has to be replaced with:
\begin{equation}
\begin{split}
I_{i+1}^{\mathrm{new}}(\mathrm{in},p) &= I_{i+2}^{\mathrm{old}}(\mathrm{in},p) e^{-\Delta \tau_{i+2,i+1}^{\mathrm{old}}}  \\
&+ \Psi_{i+2}^{\mathrm{new}}(\mathrm{in}) S_{i+2}^{\mathrm{old}}(p) +\Psi_{i+1}^{\mathrm{new}}(\mathrm{in}) S_{i+1}^{\mathrm{old}}(p) + \Psi_{i}^{\mathrm{new}}(\mathrm{in}) S_{i}^{\mathrm{new}}(p) ,
\end{split}
\end{equation}
since the populations at point $i$ have just been corrected. This correction in the specific intensity is performed for each ray
and is then taken into account when calculating the mean intensity at point i+1.

\begin{figure}
\centering
\includegraphics[width=5cm]{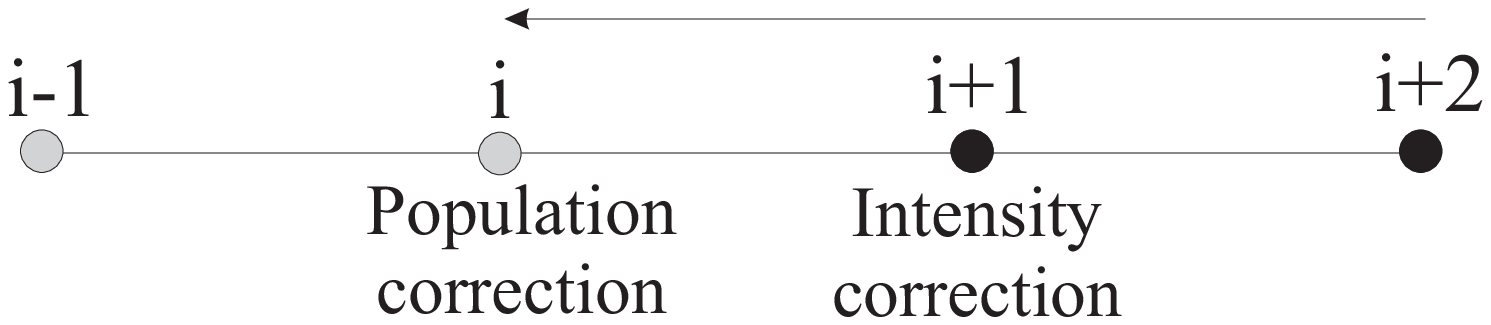}\hspace{1cm}%
\raisebox{-1cm}{\includegraphics[width=5cm]{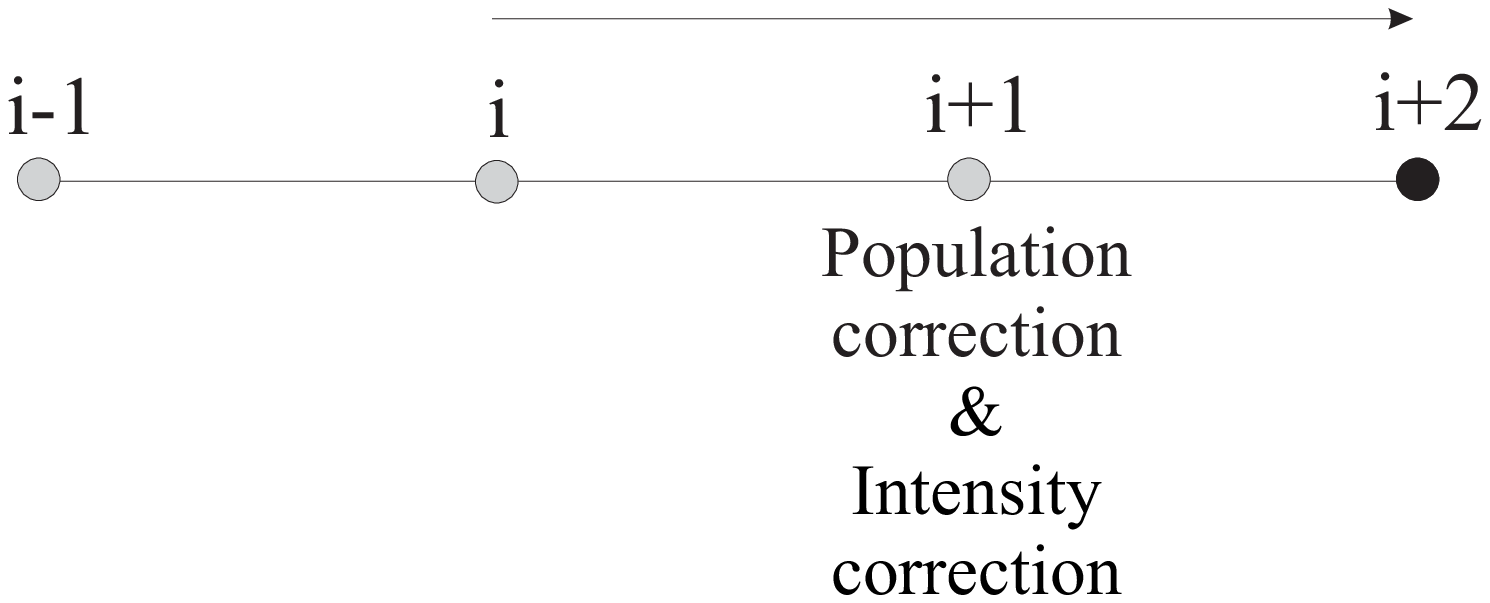}}
\caption{Corrections needed for obtaining a true Gauss-Seidel iteration when using the parabolic SC method. The left panel shows the corrections to the incoming
radiation field at point $i+1$ due to the population correction obtained previously at point $i$. The right panel shows the corrections to the
outgoing radiation field at point $i+1$ after performing the population correction at this point $i+1$. Note that the inner boundary is at point $i=1$ and
the outer boundary is at point $i=\mathrm{NP}$.}
\label{fig_GS_corrections}
\end{figure}

\item[b)] The second correction, represented in the right panel of Fig. \ref{fig_GS_corrections}, is necessary once a new estimation of the populations is
available
for point $i+1$. Since the outgoing radiation field at point $i+1$ has been obtained with the ``old'' populations at this point, it
has to be corrected in order to obtain the true value of the outgoing radiation field at point $i+1$ before moving to the next point $i+2$.
Therefore, the outgoing intensity at point $i+1$, which had been obtained using:
\begin{equation}
\begin{split}
I_{i+1}^{\mathrm{old}}(\mathrm{out},p) &= I_{i}^{\mathrm{new}}(\mathrm{out},p) e^{-\Delta \tau_{i,i+1}^{\mathrm{new}}} \\
&+ \Psi_{i}^{\mathrm{new}}(\mathrm{out}) S_{i}^{\mathrm{new}}(p) +
\Psi_{i+1}^{\mathrm{new}}(\mathrm{out}) S_{i+1}^{\mathrm{old}}(p) + \Psi_{i+2}^{\mathrm{new}}(\mathrm{out}) S_{i+2}^{\mathrm{old}}(p),
\end{split}
\end{equation}
has to be replaced by:
\begin{equation}
\begin{split}
I_{i+1}^{\mathrm{new}}(\mathrm{out},p) &= I_{i}^{\mathrm{new}}(\mathrm{out},p) e^{-\Delta \tau_{i,i+1}^{\mathrm{new}}} \\
&+ \Psi_{i}^{\mathrm{new}}(\mathrm{out}) S_{i}^{\mathrm{new}}(p) +
\Psi_{i+1}^{\mathrm{new}}(\mathrm{out}) S_{i+1}^{\mathrm{new}}(p) + \Psi_{i+2}^{\mathrm{new}}(\mathrm{out}) S_{i+2}^{\mathrm{old}}(p)  ,
\end{split}
\end{equation}
since now the populations at point $i+1$ have just been corrected. 
\end{enumerate}
These corrections are applied consecutively while computing the outgoing radiation field until reaching the outer boundary surface.

Summarizing, when such two corrections are applied to the iterative scheme, a true Gauss-Seidel iteration is obtained once the 
specific intensity of the radiation field is propagated from the outer to the inner boundary, and again to the 
outer one. This MUGA scheme leads to a convergence rate a factor 2 higher than that corresponding to the MALI method. \emph{The main advantage of this GS-based iterative scheme is that 
it has the convergence properties of a triangular $\mathbf{\Lambda}_{\nu \mathbf{\Omega}}^*$-operator, but neither the construction nor the inversion of this 
approximate operator has to
be performed}. Therefore, it is also suitable for 2D and 3D multilevel radiative transfer (see Fabiani Bendicho \& Trujillo Bueno 1999).

As pointed out by Trujillo Bueno \& Fabiani Bendicho (1995) and by Trujillo Bueno (2003), the increase in the convergence rate can be 
enhanced to a factor of 4 if the previous corrections are applied, not only when
the radiation field is being propagated outwards, but also when it is being propagated inwards. In this way, each incoming+outgoing pass 
produces two GS iterations. Some examples of 
the application of this ``up+down'' strategy to polarized radiative transfer can be found
in Trujillo Bueno \& Manso Sainz (1999).

\subsubsection{The Multilevel Successive Overrelaxation Method (MUSOR)}
The Successive Overrelaxation method (SOR) has been also proposed by Trujillo Bueno \& Fabiani Bendicho (1995) for the solution of non-LTE RT problems. The solution process is exactly the same as in MUGA, since the only difference lies on the level population correction. In the SOR method, a parameter $\omega$ is introduced to overcorrect the population, thus allowing for some kind of anticipation of the future corrections. This scheme can be written as
\begin{eqnarray}
\label{eq_SOR_method}
\bf A_{\rm GS}^{\rm old \& new} \cdot \delta n & = & \bf r \nonumber \\
\bf n^{\rm new} & = & \bf n^{\rm old} + \omega \delta n,
\end{eqnarray}
where the $\omega$ parameter has a value between 1 and 2 in order to produce the overcorrection. The MUGA method is therefore a 
particular case of the MUSOR scheme (see Trujillo Bueno \& Fabiani Bendicho 1995 for understanding how to calculate the optimum value of the parameter $\omega$).

\subsection{Computational time}
A fundamental consideration of any iterative method is to know how the total computational time scales with the number of spatial
grid points, rays, frequencies and radiative transitions in the chosen atomic or molecular model. Let $N_\mathrm{atm}$ be the number of points in the atmosphere, $N_{\mu}$ the number of points of the angular quadrature for computing the mean intensity, 
$N_\nu$ the number of frequencies at which the RT equation has to be solved and $N_\mathrm{iter}$ the number of iterations needed for
reaching
the convergence in the chosen spatial grid. With the short-characteristics method, the computational
time $T_\mathrm{cpu}$ scales linearly with the number of spatial grid points in the atmosphere, the number of frequency points and the number of
angles, so that:
\begin{equation}
T_\mathrm{cpu} \propto N_\nu N_\mu N_\mathrm{atm} N_\mathrm{iter}.
\end{equation}
The dependence of the total computational time on the number of points in the atmosphere scales as $N_\mathrm{atm}^2$ for MALI, as
$N_\mathrm{atm}^2/2$ for MUGA and as $\sqrt{N_\mathrm{atm}} N_\mathrm{atm}/2$ for MUSOR (Trujillo Bueno \& Fabiani Bendicho 1995). This is a direct consequence of the different
number of iterations required for obtaining convergence with each of the methods in the chosen spatial grid: it is of the order of $N_\mathrm{atm}$
for MALI, $N_\mathrm{atm}/2$ for MUGA and $\sqrt{N_\mathrm{atm}}$ for MUSOR. As pointed out by Trujillo Bueno (2003) the computational time of MUGA and MUSOR can be reduced further via the application of the above mentioned ``up+down'' strategy.

\subsection{Illustrative examples}
Some interesting examples of the application of the MUGA method to three-dimensional (3D) multilevel radiative transfer
can be seen in Fabiani Bendicho \& Trujillo Bueno (1999). Model calculations of scattering polarization signals in spectral
lines can be found in Trujillo Bueno \& Landi Degl'Innocenti (1997), Trujillo Bueno \& Manso Sainz (1999), Manso Sainz \& Trujillo Bueno (2003) and Trujillo Bueno (2003). Detailed information on the convergence properties of the MUGA method in one-dimensional
plane-parallel atmospheres
can be found in Fabiani Bendicho, Trujillo Bueno \& Auer (1997).
Here we show some illustrative examples of our generalization of the MUGA and MUSOR methods to multilevel RT in spherical
geometry.

\subsubsection{The limiting case of a plane--parallel atmosphere}
\label{eq_quasi_plane_parallel}
In order to show the convergence properties of these methods in spherical symmetry, we have selected the
multilevel problem treated by Avrett \& Loeser (1987), which concerns a simplified
three-level hydrogen atomic model without continuum in an isothermal semi-infinite atmosphere with
T=5000~K. The atmosphere is $\sim$1500~km thick and the radius of the internal core is 6.95$\times$10$^{10}$ cm. This may be
considered as an isothermal
representation of the solar atmosphere in spherical geometry.
The collisional rates among the three levels of the model atom are assumed to be constant. 

We define the curvature $q$ of a given stellar atmosphere as the ratio between its extension ($\Delta R$) and the radius of the star ($R$):
\begin{equation}
\label{eq_curvature}
q = \frac{\Delta R}{R},
\end{equation}
so that the RT problem should be solved using spherically symmetric geometry when $q \gtrsim 1$, while using 
plane--parallel geometry when $q \ll 1$. The case under study has $q \approx 0.002$ and may be safely solved using the plane--parallel approximation.

\begin{figure}
\centering
\includegraphics[width=0.45\textwidth]{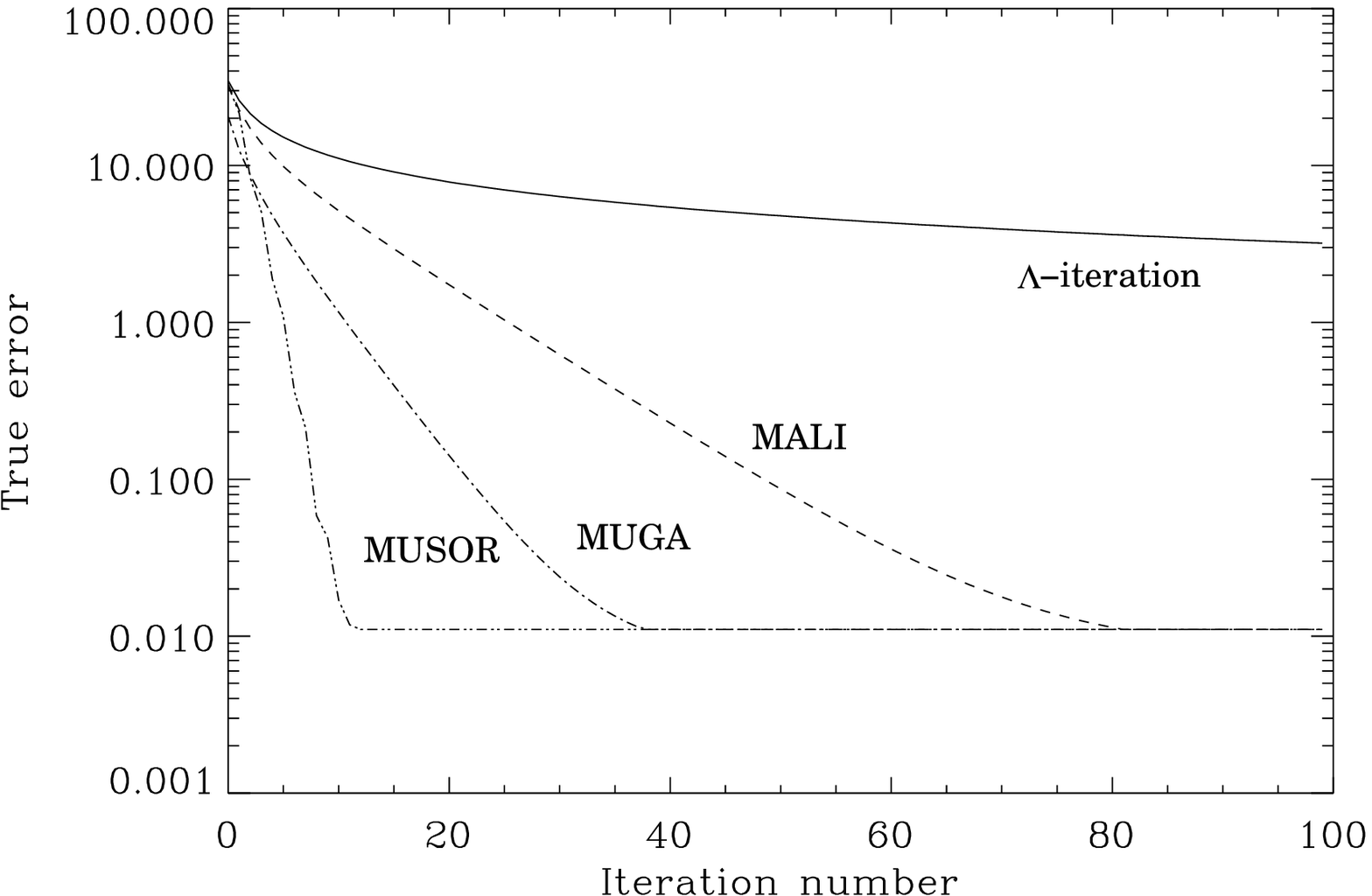}\hspace{1cm}%
\includegraphics[width=0.45\textwidth]{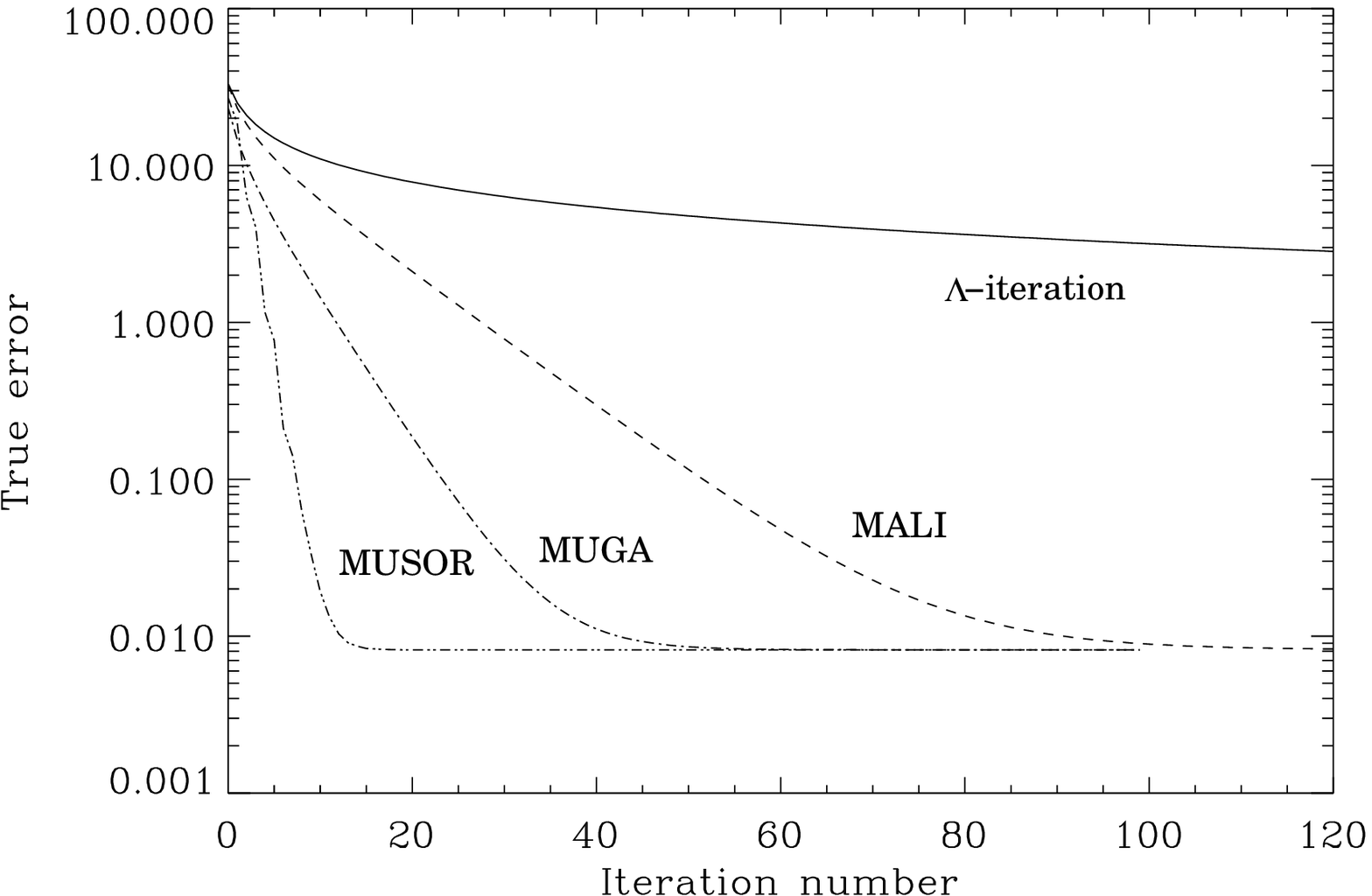}
\caption{Convergence properties of the various iterative methods showing the variation of the true error versus the iteration number for a 
three-level hydrogen atomic model. The left
panel shows the results for the quasi plane--parallel problem ($q \approx 0.002$) using our spherical geometry MUGA code while the right panel shows the same result but using the plane--parallel version.}
\label{fig_hydrogen_true_error}
\end{figure}

We have solved this non-LTE problem with the following iterative methods: $\Lambda$-iteration, MALI, MUGA and MUSOR. Both the
spherical and plane-parallel MUGA/MUSOR radiative transfer codes have been used. In the left panel of Fig. \ref{fig_hydrogen_true_error} we show the
convergence behavior for the spherical case. In order to calculate the true error we have obtained first the converged solution in
a grid of 150 radial shells with a grid step of $\Delta r=$10 km, sampling the $\sim$1500 km thick atmosphere. The problem was then
solved using a coarser grid with 80 radial shells ($\Delta r=$ 19 km). Note that the
true error corresponding to this grid is approximately 1~\%.
The $\Lambda$-iteration method, although converging, is not capable of finding the solution in a suitable number
of iterations due to the large optical depth of the transitions in the model atmosphere.

Concerning the results obtained with our MUGA codes, the convergence is reached with half the number of iterations required by the MALI method, which leads to approximately half of the total computational time, since the computational time per iteration is
approximately the same in both methods. Finally, as shown in Fig. \ref{fig_hydrogen_true_error} the number of iterations required by
the MUSOR method is a factor 10 smaller.

The convergence behavior when the same problem is solved using the plane-parallel approximation is shown in the right panel of Fig.
\ref{fig_hydrogen_true_error}. It is interesting to note that, due to the low curvature radius of the atmosphere,
both results are very similar. However, the total computational time is larger when the problem is solved with our spherically symmetric 
code due to the large number of formal solutions that have to be performed along rays of constant impact parameter.

\subsubsection{Spherical case}
\label{eq_spherical_case}
Let us consider now a problem which is very similar to the previous one, but with a more extended atmosphere.
The core radius is still 6.95$\times$10$^{10}$~cm, but the total thickness of the atmosphere is now increased to
6.3$\times$10$^6$~km, which implies a curvature of $q \sim 9$. With this $q$--value, sphericity effects have a great influence
on the final result because there are now directions in
the atmosphere along which the photons can escape more easily. The temperature of the atmosphere is still T=5000~K. We have chosen a
different non-LTEmultilevel problem characterized by a simplified 5--level Ca {\sc ii} model atom with
constant collisional rates (Avrett \& Loeser 1987). 

\begin{figure}
\centering
\includegraphics[width=0.45\textwidth]{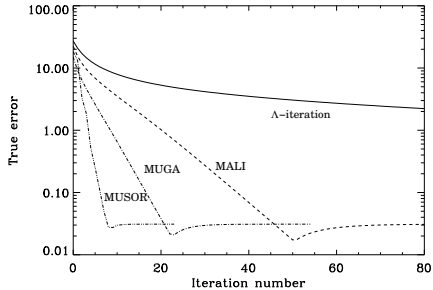}\hspace{1cm}%
\caption{Convergence properties of the various iterative methods showing the true error versus the iteration number for a multilevel 
RT problem for which sphericity effects are very important. In this case, we used a Ca {\sc ii} model with 5 levels.}
\label{fig_calcium_true_error}
\end{figure}

In order to be able to compute the true error at each iterative step as indicative of the convergence properties, the fully converged solution
has been obtained first in an atmosphere sampled at 300 radial shells. The atmosphere is now $\sim 4000$ times more extended. Therefore,
in order to obtain a reasonable precision, the number of radial shells have to be greater than in the quasi plane--parallel case of the
previous section. Figure \ref{fig_calcium_true_error} shows the true error
for the four iterative schemes. The calculations have been performed using
a grid of 150 radial shells. The convergence behavior is very similar to the previous case, but the true error of the final
converged solution is 3~\% due to
the larger distance between grid points. As expected, the
$\Lambda$-iteration method does not reach convergence in a suitable number of iterations (the optical depths are well above 10$^6$). The convergence properties of the rest of the methods
is similar to that of the previous quasi plane--parallel calculation. If $N$ is the number of iterations that MALI needs to reach convergence, MUGA reaches convergence in roughly $N/2$ iterations while with our MUSOR code convergence is reached after only $\sqrt{N}$ 
iterations. 

\subsubsection{A realistic multilevel problem: warm water in SgrB2}
\label{eq_warm_water}
\begin{figure}
\centering
\includegraphics[width=0.45\textwidth]{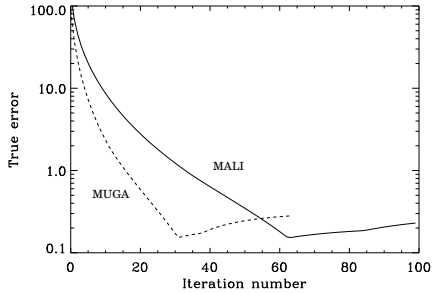}\hspace{1cm}%
\caption{Convergence properties of the MALI and MUGA methods when solving the multilevel water problem described in the text. The time 
needed to reach the
final true error with the MALI method is two times larger than with MUGA.}
\label{fig_sgrb2}
\end{figure}

Finally, we investigate the convergence properties of the MALI and MUGA methods for the solution of a complicated
multilevel radiative transfer problem in spherical geometry. This problem consists in the calculation of the self-consistent populations
of the first 14 rotational levels of ortho-water in a warm ($T\sim 300-500$ K) shell. This scenario tries to model the physical conditions under which many 
H$_2$O lines observed with ISO are formed (Cernicharo et al. 2006). Water introduces some difficulties in the convergence 
process because of the appearance of maser transitions (specially the line at 183 GHz). The convergence properties are shown in Fig. \ref{fig_sgrb2}
when the model atmosphere is discretized with 150 shells. Convergence with the MUGA method
is reached with half the number of iterations required by the MALI code, similar to the previous illustrative examples. The efficiency can be
improved further via the ``up+down'' strategy proposed by Trujillo Bueno \& Fabiani Bendicho (1995) and/or via MUSOR and/or 
via the application of acceleration techniques.

\section{Molecular Number Densities and Non-equilibrium Chemistry}
Molecules are usually found in highly dynamic systems (e.g., the solar atmosphere, winds of AGB stars, etc.) and their formation
is influenced by the time variation of the physical conditions in the medium. Therefore, only when the dynamical timescales are
much slower than the timescales of molecular formation, the approximation of Instantaneous Chemical Equilibrium (ICE) can
be used safely. Under the ICE approximation, molecules are assumed to be formed instantaneously and
their abundances depend on just the local temperature and density. Another consequence of this assumption is that the specific
reaction mechanisms that create and destroy a given molecule are irrelevant and only the molecule and its
constituents are important. The ICE approximation has been applied to many astrophysical
situations (Russell 1934; Tsuji 1964; Tsuji 1973; McCabe, Connon Smith \& Clegg 1979, etc.).

\subsection{Basic Equations}
When the dynamical timescales are smaller than the time needed to reach the molecular equilibrium
number densities, it is fundamental to consider all the reactions which create and destroy a given species, which requires solving the full
chemical evolution (CE) problem. The temporal variation of the abundance of a given species $i$ can be modeled with the following set of
nonlinear ordinary differential equations (see, e.g., Bennett 1988):
\begin{eqnarray}
\label{eq_chem_sys}
\frac{dn_i}{dt} &=& \sum_A \sum_B \sum_C k_{ABC} n_A n_B n_C + \sum_A \sum_B k_{AB} n_A n_B  + \sum_A k_{A} n_A \\
&- & \sum_A \sum_B k_{ABi} n_A n_B n_i  - \sum_A k_{Ai} n_A n_i - k_i n_i \nonumber.
\end{eqnarray}
The set of reactions is classified according to the number of species which have to collide to let the reaction take place:
\begin{itemize}
\item \textbf{Three-body reactions}. The reactions are of the type $\mathrm{A}+\mathrm{B}+\mathrm{C} \to products$. They are
included in the model by means of the first and the fourth terms of Eq. (\ref{eq_chem_sys}). They correspond to reactions which create
and destroy species $i$ from the reaction of other three components, respectively. They are
characterized by the
reaction rates $k_{ABC}$ in units of cm$^6$ s$^{-1}$ whose value is usually extremely small. They typically give a negligible
contribution to the total rate of variation of the abundance of species $i$ for low density media.
On the other hand, for relatively high density media like the
photospheric plasma of solar-like atmospheres these reactions become extremely important. For example, the catalytic formation of hydrogen by
the three body reaction H+H+H $\to$ H$_2$+H has a rate of $\sim$10$^{-30}$ cm$^6$ s$^{-1}$. Then, since the typical hydrogen densities
in the solar photosphere are of the order of 10$^{15}$-10$^{17}$ cm$^{-3}$, the rate of formation of H$_2$ turns
out to be of the order of 10$^{15}$-10$^{21}$ cm$^{-3}$ s$^{-1}$.

\item \textbf{Two-body reactions}. They are of the form $\mathrm{A}+\mathrm{B} \to products$, and are included in the
model by means of the second and fifth terms in Eq. (\ref{eq_chem_sys}). They correspond to reactions which create and destroy species $i$
from the reaction of other two components, respectively. They are
characterized by the reaction rates $k_{AB}$ in units of cm$^3$ s$^{-1}$ and typically represent the most important reactions in
almost all the astrophysical media in which chemical reactions take place.

\item \textbf{One-body reactions}. These reactions are of the form $\mathrm{A} \to products$. They usually represent
photodissociation or photoionization in which the species is dissociated by photons of energy greater than the dissociation
potential (this is only valid for molecular species because atomic species do not dissociate) or ionized by photons higher than
the ionization potential (valid both for atomic or molecular species). Therefore, these reactions are often written as
$\mathrm{A} + h\nu \to products$, where we have explicitly indicated the necessity of a radiation field for the reaction to take place.
This dependence on the
radiation field present in the medium is implicitly taken into account in the reaction rate $k_{A}$ in units of s$^{-1}$.
The model includes such reactions by means of the third and sixth terms in Eq. (\ref{eq_chem_sys}), corresponding to those which
create and destroy species $i$, respectively.
\end{itemize}

Once the physical conditions, the set of species included in the model, and the reaction rates of the reactions which take place
among them are known, the system of equations given by Eq. (\ref{eq_chem_sys}) is completely defined. This set of nonlinear differential
equations represents a very \emph{stiff} problem because of the different variables and rates having disparate ranges of variation.
A suitable method that can cope with this stiffness has to be used. For instance, an algorithm based on the backward differentiation
formula (see, e.g., Gear 1971) can assure stability and allows us to use large time steps so that the solution for
very long evolution times can be obtained. An interesting application which helped to solve the enigma of CO ``clouds'' in the solar
atmosphere can be seen in Asensio Ramos et al. (2003).

\subsection{Relaxation times}
\label{sec_relaxation_times}
In a medium of given temperature $T$ and hydrogen number density $n_\mathrm{H}$, it is of interest
to ask for the formation time of a given molecule when the physical conditions are perturbed. If the temperature and density values permit
fast reactions, the time for molecular formation will be short. If reactions take place very slowly, molecular number densities will 
adapt to the new conditions after a very long timescale. Therefore, short \emph{relaxation times} may be interpreted as an indication
that the ICE approximation will give a correct description of the molecular number densities, while large relaxation times would
indicate that we better take into account the time evolution of the molecular number densities.

The calculation is initialized assuming that at $t=0$ s we only have atomic species. Then, the system evolves until
reaching the chemical equilibrium abundances. These molecular number densities should be similar to those obtained via the ICE approximation. 
For this to be true, the set of reactions and the chemical species included in the model have to be complete enough.
Once such equilibrium abundances are obtained, we introduce a perturbation in the physical conditions and we calculate the time needed to reach the 
new equilibrium situation compatible with the new physical conditions. This timescale is called the \emph{relaxation time}.
We assume that the perturbations are small enough
to consider that we are in the linear regime (i.e., $|\Delta T|/T \ll 1$, $|\Delta n_\mathrm{H}|/n_\mathrm{H} \ll 1$).

\begin{figure}
\centering
\includegraphics[width=0.45\textwidth]{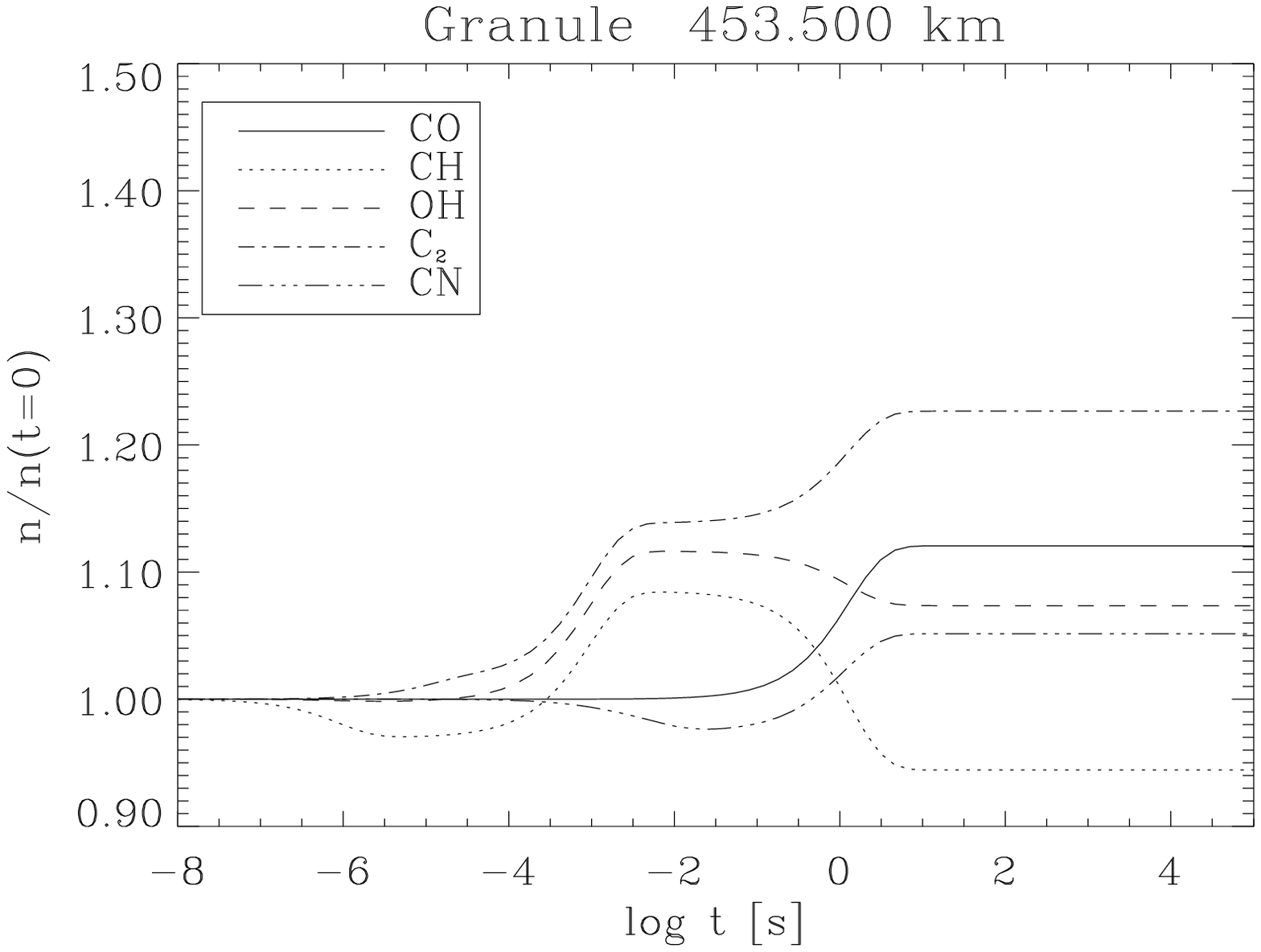}%
\includegraphics[width=0.45\textwidth]{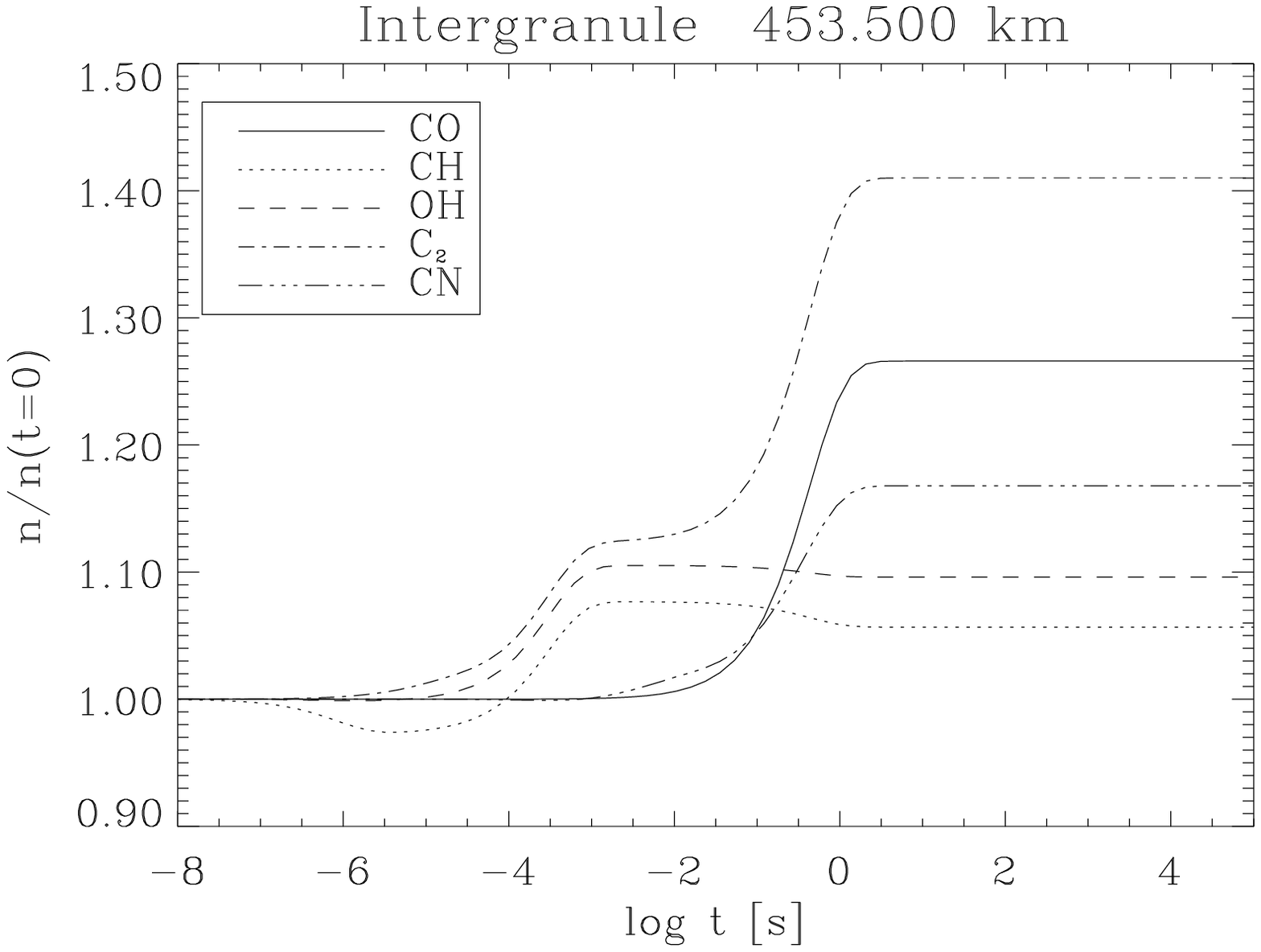}
\caption{Time evolution of the molecular number densities for some selected molecules at some heights of the granular and intergranular
models and the semi-empirical models FAL-C and COOL-C. At $t=0$ s, a reduction of 1\% in the temperature has been performed
and the evolution until reaching the new equilibrium
abundance has been obtained. Note the simple behavior of the CO abundance, while the rest of molecules have a complicated
evolution showing the extremely non-linear character of the chemical evolution problem. All the molecules reach their new equilibrium
concentrations in less than 10 s, except for the points at $\sim$700 km in both semi-empirical models. Since the granular and intergranular
models represent only the photospheric conditions of the solar atmosphere, no high relaxation times are found.}
\label{fig_evolution_after_perturbation}
\end{figure}

We have selected a granular and an intergranular model from the 3D hydrodynamical simulations of solar surface
convection carried out by Asplund et al. (2000). While the granular model is hotter in the deep regions of the photosphere, it turns out to be cooler
above $\sim$200 km. Figure \ref{fig_evolution_after_perturbation} shows the molecular number density relative to that at 
$t=0$ s for some selected
species at a height of 453 km. In this calculation, the temperature has been reduced by 1\%. It is clearly seen that almost all 
the chemical species increase their abundance after the perturbation, once the new equilibrium situation is reached. The relaxation times in the
granular plasma are larger than in the intergranular gas because the temperature and hydrogen number density is smaller at such heights.
Both the lower temperature and hydrogen number density produce a lower rate of collisions which results in slower reactions

As seen in Fig. \ref{fig_evolution_after_perturbation}, the time evolution of the molecular number densities seems to be quite simple for the CO molecule, which follows a smoothed step
function. On the other hand, it is more complicated for the majority of the molecules, with episodes in which the abundance is increasing and
others in which it is decreasing. The quite simple time variation of the CO abundance can be explained with a very simplified 
model similar to that used by Ayres \& Rabin (1996). Since CO is one of the most
abundant molecules in the solar atmospheric environment, we can assume that its formation and destruction depends on two processes. On the one
hand, direct association of C and O to form CO:
\begin{equation}
\label{reac_CO}
\begin{split}
\mathrm{C} + \mathrm{O} & \xrightarrow{K_{\mathrm{CO}}} \mathrm{CO}, \\
\mathrm{CO} & \xrightarrow{K'_{\mathrm{CO}}} \mathrm{C} + \mathrm{O},
\end{split}
\end{equation}
where $K_{\mathrm{CO}}$ and $K'_{\mathrm{CO}}$ are the reaction rate for the formation of CO by neutral association of
its constituents and the reaction rate for the dissociation of CO, respectively. 
Another reaction path which efficiently forms CO is:
\begin{equation}
\begin{split}
\mathrm{C} + \mathrm{OH} & \xrightarrow{K_1} \mathrm{CO} + \mathrm{H} \\
\mathrm{CO} + \mathrm{H} & \xrightarrow{K_2} \mathrm{OH} + \mathrm{C}.
\end{split}
\end{equation}
It can be verified that the last pair of chemical reactions are much more efficient in setting the CO number density than those
given by Eqs. (\ref{reac_CO}). 
As a first approximation, we can assume that OH is formed much faster than CO and so its abundance can be considered constant
during the time interval in which the number density of
CO molecules is changing. This fact is reinforced by our chemical evolution calculations, as seen in
Fig. \ref{fig_evolution_after_perturbation}. Note that OH has almost reached its equilibrium abundance when CO starts to
change from its value at $t=0$ s.
Consider now the simplifying condition that the hydrogen number density is constant. The OH number density can be written in terms of
that of hydrogen 
as $n_{\mathrm{OH}} \approx A_{\mathrm{OH}} n_{\mathrm{H}}$ (where $A_{\mathrm{OH}}$ is the OH abundance relative to hydrogen),
while the carbon abundance is given by $n_{\mathrm{C}} \approx A_{\mathrm{C}} n_{\mathrm{H}}-n_{\mathrm{CO}}$ (where $A_{\mathrm{C}}$ is
the carbon abundance relative to hydrogen). The first assumption introduces a very small error since the hydrogen number density is
barely affected by the presence of molecular species unless the temperature is extremely small. The second assumption can be 
considered suitable since $n_\mathrm{OH}$ reaches its equilibrium value much faster than $n_\mathrm{CO}$. The third
assumption is applicable because almost all the carbon not in atomic form is indeed in the form of CO.
Therefore, the differential equation that describes the time evolution of the CO number density can be rewritten as:
\begin{equation}
\frac{dn_{\mathrm{CO}}}{dt} = K_1 \left (A_{\mathrm{C}} n_{\mathrm{H}} - n_{\mathrm{CO}} \right) n_{\mathrm{OH}} -
K_2 n_{\mathrm{CO}} n_{\mathrm{H}}.
\end{equation}
The solution to this simple equation is
\begin{equation}
\label{eq_soluc_CO1}
n_{\mathrm{CO}}(t) = n_{\mathrm{CO}}(\infty) + \left[ n_{\mathrm{CO}}(0)-n_{\mathrm{CO}}(\infty) \right] e^{-t/\tau_\mathrm{relax}},
\end{equation}
where
\begin{equation}
n_{\mathrm{CO}}(\infty) = \frac{K_1 A_\mathrm{OH} A_\mathrm{C} n_\mathrm{H}}{K_1 A_\mathrm{OH} + K_2},
\end{equation}
and
\begin{equation}
\tau_{\mathrm{relax}} = \frac{1}{\left( K_1 A_{\mathrm{OH}} + K_2 \right) n_\mathrm{H}}.
\end{equation}

Our calculations show that, if the behavior of the molecular number density can be written in a simple way like for
that of the CO case, its evolution can be characterized by a timescale which depends on the main reactions which create and destroy the
molecule and by its equilibrium abundance (which also depend on the main reactions). Note that it is not correct to define
timescales for the formation and destruction processes separately, since there is only one global timescale which involves both
processes. This is the reason that explains why one cannot know how fast a chemical reaction is going to
take place by only taking into account the rate of the individual reaction. When the
whole system of differential equations is solved, many non-linearities arise, making it very difficult to make a simple 
relaxation time estimation.

\begin{figure}
\centering
\includegraphics[width=10cm]{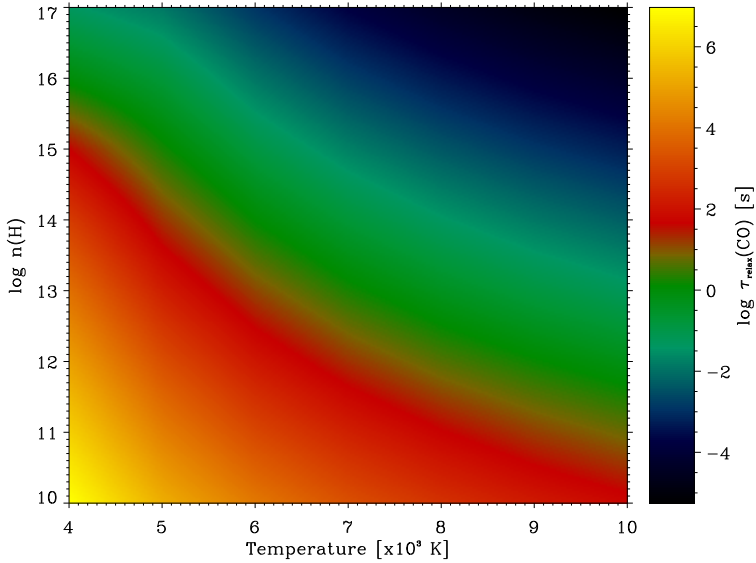}
\caption{Relaxation time of the CO number density
for a broad range of possible physical conditions in a stellar atmosphere. Note
that the relaxation time increases when the temperature and/or hydrogen number
density are reduced, while it decreases when the temperature and/or hydrogen number density are increased.}
\label{fig_CO_relaxation_time}
\end{figure}

The linear analysis of the time evolution of the CO number density shows that CO is mainly driven by itself as dictated by Eq.
(\ref{eq_soluc_CO1}). Therefore, it is possible to carry out a fit of this temporal evolution behavior 
and obtain a relaxation time $\tau_\mathrm{relax}$. Since a linear analysis is suitable for CO, the
relaxation time can be calculated for a range of perturbations in the temperature and hydrogen number density. The results are shown
in Fig. \ref{fig_CO_relaxation_time}. The general trend is that, \emph{for a given density, the relaxation time
is the larger the cooler the medium where the temperature perturbation is introduced}. Similarly, \emph{for a given temperature
of the unperturbed medium, the relaxation time increases rapidly with decreasing density}. Short relaxation
times are typical of high-temperature and high-density media (e.g. $t_{\rm relax}\,{\approx}\,0.006$ s for
$n_{\rm H}=10^{16}\,{\rm cm}^{-3}$ and $T=6000$ K), while long relaxation times are characteristic of low-temperature
and low-density situations (e.g. $t_{\rm relax}\,{\approx}\,400$ s for $n_{\rm H}=10^{14}\,{\rm cm}^{-3}$
and $T=4000$ K). This behavior can be easily understood in terms of the number of collisions per unit time, which decreases both when
the temperature and/or the density decreases. In a realistic situation like in the solar atmosphere, one would find a broad range of 
relaxation times at a fixed atmospheric height because the existing rarefactions, compressions and temperature fluctuations
are continually changing the physical conditions. Obviously, the situation is highly non-linear and 
the relaxation time concept, although useful,
loses its meaning in the real Sun. Therefore, any firm conclusion needs to be achieved via detailed numerical simulations
(see Asensio Ramos et al. 2003).

\section{Concluding remarks}

Thirty years ago, the total computational work ($W$) of the best available multilevel transfer codes scaled approximately as 
NP$^3$, with NP the number of spatial grid points in a computational domain of fixed dimensions (e.g., Mihalas 1978). Twenty years ago,
 operator splitting methods based on Jacobi iteration were developed that yielded $W{\sim}{\rm NP}^2$ (e.g., Olson, Auer \& 
Buchler 1986; Rybicki \& Hummer 1991, 1992; Auer, Fabiani Bendicho \& Trujillo Bueno 1994). Ten years ago, radiative transfer methods based on Gauss-Seidel and SOR iteration 
were developed for which $W{\sim}{\rm NP}{\,}\sqrt{{\rm NP}}$, which imply an-order-of-magnitude of improvement with respect to 
the MALI method (Trujillo Bueno \& Fabiani Bendicho 1995; Trujillo Bueno \& Manso Sainz 1999; Trujillo Bueno 2003).
Such methods are also suitable for 2D and 3D radiative transfer applications because they do not require neither the construction
nor the inversion of any non-local approximate $\Lambda$-operator (see the MUGA-3D code of 
Fabiani Bendicho \& Trujillo Bueno 1999). Moreover, MUGA is the iterative method of choice for the development of the only multilevel RT method for which $W{\sim}{\rm NP}$ -that is, the non-linear multigrid method described by Fabiani Bendicho, Trujillo Bueno \& Auer (1997).

In this article we have shown how the Multilevel Gauss-Seidel (MUGA) and Multilevel Successive Overrelaxation (MUSOR) methods can
be applied for solving multilevel radiative transfer problems in spherical geometry, for both atomic and molecular lines. Our multilevel 
codes based on the MUGA and MUSOR methods offer a very powerful tool for the fast solution of realistic RT problems.

\acknowledgements
We thank Philippe Stee for his kind invitation to present this keynote paper in a highly interesting workshop.
This research has been funded by the European Commission through the Solar Magnetism Network (contract
HPRN-CT-2002-00313) and by the Spanish Ministerio de Educaci\'on y Ciencia through project AYA2004-05792.


\end{document}